\documentclass{svjour3}                     % onecolumn (standard format)
\smartqed  % flush right qed marks, e.g. at end of proof
\usepackage{graphicx}
\usepackage{mathptmx}      % use Times fonts if available on your TeX system
\usepackage{natbib}
\def\kms{km~s$^{-1}$}
\def\ga{\mathrel{\hbox{\rlap{\hbox{\lower4pt\hbox{$\sim$}}}\hbox{$>$}}}}
\def\la{\mathrel{\hbox{\rlap{\hbox{\lower4pt\hbox{$\sim$}}}\hbox{$<$}}}}
\journalname{SSRv}
\begin{document}

\title{Exotic clouds in the local interstellar medium}
%\subtitle{Exotic clouds in the local interstellar medium}


\author{Sne\v{z}ana Stanimirovi\'{c}}       

\authorrunning{Stanimirovic} % if too long for running head

\institute{S. Stanimirovi\'{c} \at
              Department of Astronomy, University of Wisconsin, 475 N. Charter
              Street, Madison, WI 53706 USA  \\
              Tel.: +1-608-890-1458\\
              Fax: +1-608-263-6386\\
              \email{sstanimi@astro.wisc.edu}    }

\date{Received: date / Accepted: date}

\maketitle

\begin{abstract}

The neutral interstellar medium (ISM) inside the Local Bubble (LB) has 
been known to have properties typical of the warm neutral medium (WNM).
However, several recent neutral hydrogen (HI) absorption experiments show
evidence for the existence of at least several cold diffuse clouds  
inside or at the boundary of the LB, with properties highly unusual relative to the
traditional cold neutral medium.
These cold clouds have a low HI column density, and AU-scale sizes.
As the kinematics of cold and warm gas inside the LB are similar, this suggests
a possibility of all these different flavors of the local ISM belonging to
the same interstellar flow. 
The co-existence of warm and cold phases inside the LB is exciting as it
can be used to probe the thermal pressure inside the LB.
In addition to cold clouds, several discrete
screens of ionized scattering material are clearly located inside the LB.  

The cold exotic clouds inside the LB
are most likely long-lived, and we expect many more clouds with similar
properties to be discovered in the future
with more sensitive radio observations.
While physical mechanisms responsible for the production of such clouds are still 
poorly understood, dynamical triggering of phase conversion 
and/or interstellar turbulence are likely to play an important role.

\keywords{Interstellar medium: Physical properties \and Interstellar medium:
  Solar neighborhood}
\end{abstract}

\section{Introduction}
\label{intro}

The diffuse interstellar medium (ISM) in the Galaxy contains structure over a wide range of
spatial scales. While on scales $\ga1$ pc we can observationally trace 
the entire hierarchy of structures, the extremely small-scale end of this
spectrum, on scales from $<1$ pc to tens of AUs, is still largely unexplored. 
In terms of the diversity of physical properties, 
the diffuse neutral ISM is traditionally viewed as two types of 
``clouds'', referred to as
the Cold Neutral Medium (CNM) and the Warm Neutral Medium (WNM).
The CNM and the WNM have very different temperature and volume densities, 
however they co-exist spatially and are expected to be (from a theoretical point of
view), at least locally, in pressure equilibrium. 
The CNM has a temperature of $T\sim70$ K and a hydrogen
volume density $n\sim40$ cm$^{-3}$, while the WNM has 
$T\sim5000$ K and $n\sim0.5$ cm$^{-3}$ 
\citep{Heiles03b}. 
In terms of HI column density, the CNM has typically $5\times10^{19}$ cm$^{-2}$ (mean
value), and the WNM has $1.3\times10^{20}$ cm$^{-2}$ \citep{Heiles03b}.
The CNM clouds were traditionally assumed to be 
spherically symmetric with a size of 1-2 pc.
Interestingly, 
recent observations are finding much larger aspect ratios.

In this paper, we focus on the structure in the
diffuse ISM on spatial scales of a few tens to a few thousands of AUs. 
We show that these features have observationally inferred 
properties not in accord with the
traditional ISM picture, and therefore sample flavors of the most exotic diffuse ISM.
In particular, we discuss the tiny scale atomic and ionized structures,
and CNM clouds with very low HI column densities.
Several of these exotic features are known to be very close by, 
inside or at the boundary of the Local Bubble (LB). 
The most intriguing question is whether
these exotic clouds exist because they are related to the LB, 
or the LB has nothing to do with their existence and
physical properties.
We also explore  
how the exotic ISM compare with warm clouds within 
the LB, and
look briefly into mechanisms that could play an important 
role for the formation and survival of clouds within the Local Cavity.

\section{Tiny scale atomic structure (TSAS)}
\label{sec:first}

\begin{figure}
  \includegraphics[width=0.95\textwidth]{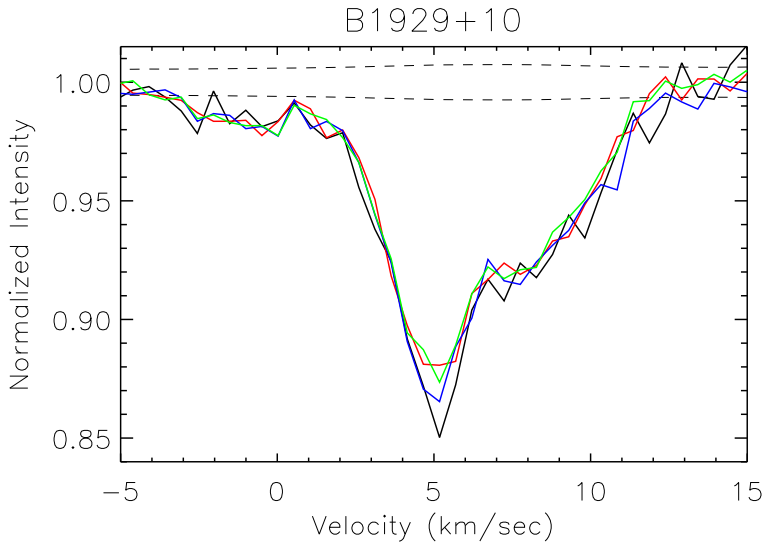}
  %{plot_change_1929_talk.ps}
\caption{HI absorption spectra in the direction of B1929+10 ($l=47^{\circ}$,
  $b=-3.9^{\circ}$) obtained with the
  Arecibo telescope at four observing epochs. Dashed lines show a typical, $\pm1-\sigma$ noise
  level in the absorption spectra. The absorption feature at 5 \kms has
  changed significantly (2-3-$\sigma$ level) with time. 
 For full analysis  please see Stanimirovic
 et al. (2008, in preparation).}
\label{fig:1}       % Give a unique label
\end{figure}

The phenomenon of Tiny Scale Atomic Structure (TSAS) was discovered in
the late 1970s when interferometric observations revealed an atomic cloud with
a size of 70 AUs \citep{Dieter76}.
When compared with traditional CNM clouds, TSAS has very exotic properties:
a typical observed neutral hydrogen (HI) column density of $10^{18-19}$ cm$^{-2}$ and a 
(typical) size of about 30 AU, imply a volume density 
$n=10^{3-4}$ cm$^{-3}$ and a thermal pressure of $P/k=nT=10^{4-6}$ cm$^{-3}$ K.
This is at least two orders of magnitude higher that what is expected for the
traditional CNM. TSAS has caused a lot of
controversy since its discovery as it obviously has properties
grossly outside of the range expected for the ISM in pressure equilibrium.
Yet, TSAS continues to be frequently encountered observationally.

How do we observe TSAS? There are several different observational methods,
one being temporal variability of HI absorption spectra against pulsars. 
Pulsars have high transverse velocities,
and typically travel over 5-50 AU per year. By comparing HI
absorption spectra obtained at two different epochs we can probe
density inhomogeneities on AU-scales in the CNM clouds that happen 
to lie between us and pulsars.
This is a powerful technique, however the difference in absorption spectra we are
after are tiny and therefore high resolution and sensitivity observations
are essential, as well as careful instrumental calibration and data reduction.

Results from the latest study of TSAS using pulsars are presented in 
Stanimirovic et al. (2008, in preparation) and \citet{Weisberg07}.  
These studies show that one pulsar in particular, B1929+10 (Figure 1),
exhibits significant changes in HI absorption spectra at several observing epochs.
The significant fluctuations, with an estimated
likelihood for being real of 90-100\% (for the full analysis
please see Stanimirovic et al. 2008, in preparation), correspond to TSAS with a typical
velocity FWHM of 1 \kms, $\Delta N(\rm HI)=10^{18}$ cm$^{-2}$, and a size 
$L=30-45$ AU. If TSAS is assumed to be spherically symmetric then, 
$n=$ a few $\times10^{3}$ cm$^{-3}$ and
thermal pressure is $P/k=$ a few $\times10^{4}$ cm$^{-3}$ K. 
However, a geometrical elongation along the line-of-sight, by a factor of $\sim10$,
would be able to bring the volume density and thermal pressure to what 
is typically expected for the CNM.

B1929+10 is the closest pulsar in our sample, at a distance of 361 $(+8, -10)$
pc \citep{Chatterjee04}. Interestingly, the line-of-sight 
toward B1929+10 runs along the elongated finger of dense neutral gas 
bounding the LB, based on Na I observations by \citet{Lallement03}.
Several studies, including \citet{Phillips92}, reported enhanced 
turbulence and scattering at the boundaries of the LB.
As half of the line-of-sight to B1929+10 is either inside or along the
wall of the Local Bubble, and also because this is the only pulsar in our sample
that shows consistently significant fluctuations, the change in HI
absorption spectra probably originates in either the presence of TSAS inside
the LB or the enhanced turbulent fluctuations along 
the wall of the LB. 

\section{Tiny scale ionized structure (TSIS)}

Analogues of TSAS exist in the ionized medium, and are called Tiny 
Scale Ionized Structures or TSIS. This phenomenon is observed by 
measuring the scattering of pulsar signals as they propagate through the 
intervening ionized medium. For excellent reviews of this phenomenon see 
\citet{Rickett07} and \citet{Stinebring07}.
In order to explain the observed scattering properties, 
two phenomena are needed in the intervening ionized medium:
(i) a smooth density distribution, with random fluctuations described with a
Kolmogorov spectrum; and
(ii) discrete structures located along the line-of-sight, with a thickness of
a few tens of parsecs (observed as ``scintillation arcs'' which were discovered 
by \citet{Stinebring01}). 
The discrete structures are often referred to as  ``thin scattering
screens''  to describe their localized and discrete nature.  
Furthermore, sometimes scattering screens have well-defined sub-structure (``arclets''), 
in the form of isolated (ionized) density fluctuations, 
with a size of about 1 AU and a volume density of about 100 cm$^{-3}$ \citep{Hill05}. 
The volume density of arclets is more than $10^3$ times larger than the average electron
density in the ISM and implies a high ionization fraction, with $T\ga10^4$ K and
$P\sim10^6$ cm$^{-3}$ \citep{Heiles07}.

Scintillation properties have been measured in directions to many pulsars and 
thin scattering screens are frequently found.
If the distance to the pulsar and its proper motion are known, 
the location of screens along the line-of-sight can be determined. 
Stinebring (2007) summarizes the locations of currently known screens, and
it is interesting to note that at least 6-7 screens are located inside the LB.
In particular, pulsars B1929+10 and B0823+26 each have 3 screens inside the
LB, while B1133+16 has a single screen at a distance of 140 pc \citep{Trang07}.

\section{Low column density clouds}

\begin{figure}
\begin{center} 
 \includegraphics[width=0.8\textwidth]{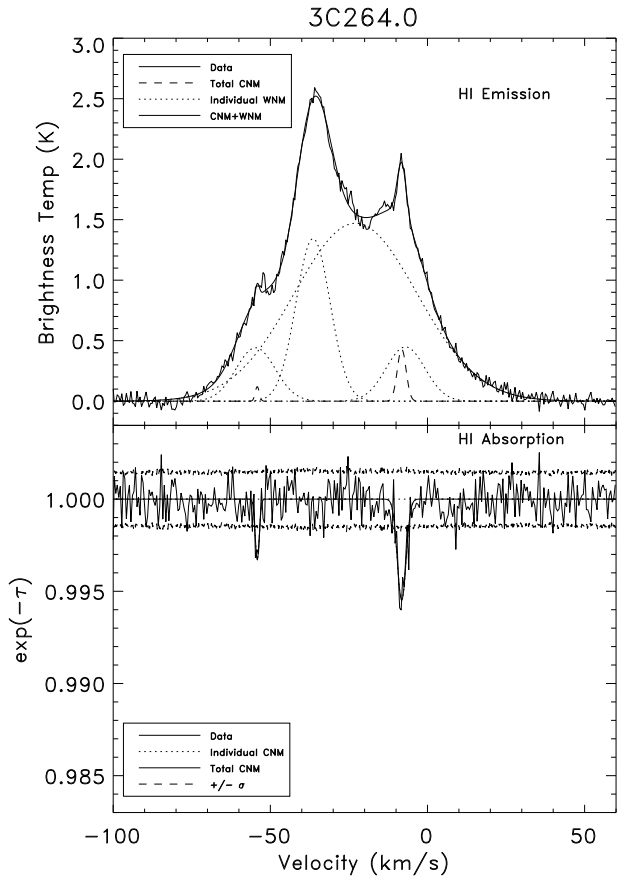}
\caption{HI emission and absorption spectra in the direction of 3C264.0, obtained with the
  Arecibo telescope (Stanimirovic, De Back et al., in preparation). 
Two new CNM components were detected.
The top panel shows the HI emission profile with separate contributions 
from the CNM and WNM shown as dashed 
and dot-dashed lines, respectively, while
the final (simultaneous) fit is shown with the thick solid line. 
The middle panel shows the HI absorption spectrum with two fitted Gaussian
components, with dotted lines representing $\pm1-\sigma$ noise level. 
The decomposition of HI emission and absorption spectra 
into Gaussian components was performed to estimate the HI spin temperature,
employing the method of Heiles and Troland (2003). 
}
\end{center}
\label{fig:1}       % Give a unique label
\end{figure}

\begin{figure}
\begin{center} 
  \includegraphics[width=0.8\textwidth]{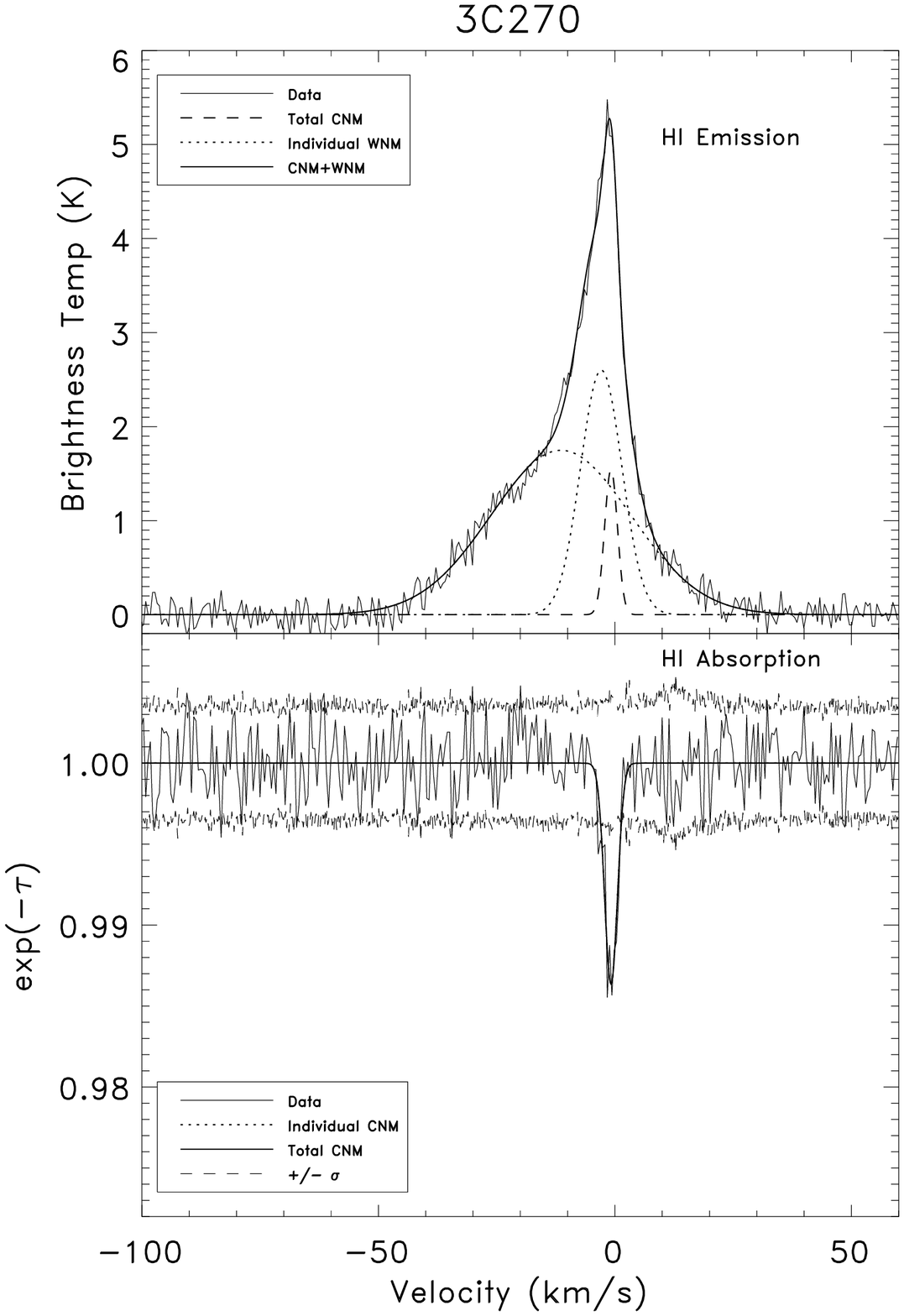}
\caption{HI emission and absorption spectra in the direction of 3C270, 
obtained with the Arecibo telescope. 
One CNM component was detected.
Plotted quantities are the same as in Figure 2.}
\end{center}
\label{fig:1}       % Give a unique label
\end{figure}

\citet{Braun05} and \citet{Stanimirovic05} recently revealed 
a new population of CNM clouds which have HI column densities about 50-100 times
lower than typical CNM clouds. We call these clouds the ``thinnest'' or
low-$N(HI)$ clouds.  Interestingly, in terms of volume density and size, 
low-$N(HI)$ clouds occupy a regime between the traditional CNM clouds and 
the extreme TSAS. The work that motivated the discovery of a large sample of 
low-$N(HI)$ clouds is the Millennium survey (Heiles \& Troland 2003), which
observed 79 continuum sources to establish temperature and column densities of
the CNM and the WNM.
Heiles \& Troland (2003) noticed that 20\% of their sight lines indicated either no 
or a very low column density of CNM. 
Stanimirovic \& Heiles (2005) and \citet{Stanimirovic07} used the Arecibo 
telescope to observe about 20 continuum sources which either had no previously
detected CNM or very weak CNM features. In total, 23 new CNM clouds were
detected, with a typical integration time of the order of a few hours per source. 
Essentially, with long integration times weak HI absorption lines emerge
easily. As an example, Figure 2 shows one of the sources, 3C264.0, where two new
CNM clouds were found with a peak optical depth of $5\times10^{-3}$ and 
$3\times10^{-3}$, or an HI column density of $9\times10^{18}$ and
$5\times10^{17}$ cm$^{-2}$, respectively. 
Clearly, these HI column densities are among the lowest ever detected for the CNM.
Another, not so extreme example is shown in Figure 3, where a single new
CNM cloud is detected at a LSR velocity of $-0.8$ \kms.

There are several important points emerging from recent studies.
(i) The detection rate of the Stanimirovic et al. (2007) experiment is very
high, suggesting that weak HI absorption features are common in the ISM. 
Essentially, we just need to integrate long enough.
(ii) The preliminary addition of the new population of CNM clouds to the
``traditional'' CNM population by Heiles \& Troland (2003) resulted in 
the combined column density histogram that can be easily fitted 
with a single function, $\propto N(HI)^{-1}$. This suggests that
low-$N(HI)$ clouds most likely do not represent a separate population, but
could simply be a low column density extension of the traditional CNM.
(iii) However, the low-$N(HI)$ clouds are probably smaller than typical CNM clouds.
If we assume an equilibrium pressure of 3000 cm$^{-3}$ K and a volume
density of about 40 cm$^{-3}$, we can estimate a cloud line-of-sight
size of about 800-4000 AU. If over-pressured, clouds would have 
even smaller line-of-sight length.
(iv) The low-$N(HI)$ clouds have an extremely low ratio of the CNM to total
HI column density, $<5$\%. This suggests that cold low-$N(HI)$
clouds are most likely surrounded by a large amount of warmer gas. This
may be important for cloud survival by providing a blanket against evaporation.

\section{Cold clouds inside the Local Bubble}

\begin{figure}
\begin{center}
  \includegraphics[width=0.8\textwidth]{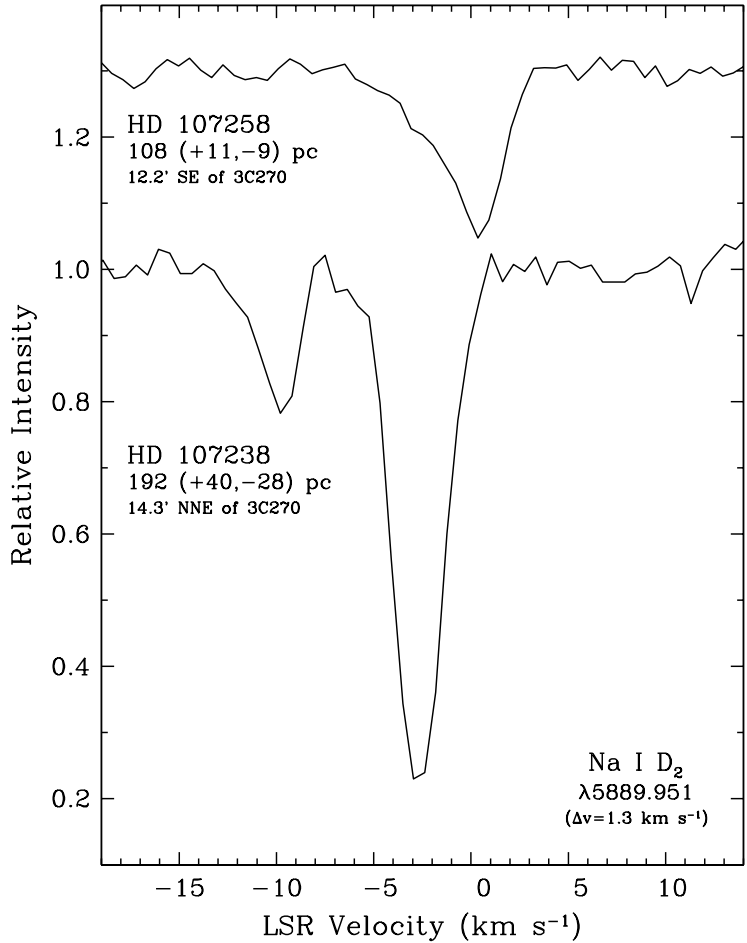}
\caption{Na I absorption spectra in the direction of 3C270.
HD 107258 is located 12 arcmin to the south-east from 3C270 and is at
a distance of 108 pc, while HD 107238 is 14 arcmin to the north-east from 3C270 and
at a distance of 190 pc. Spectra were obtained with 
the 0.9 m coud\'{e} feed telescope and spectrograph at Kitt Peak
National Observatory. }
\end{center}
\label{fig:1}       % Give a unique label
\end{figure}

For one of the low-$N(HI)$ clouds (3C270, $l=184.8^{\circ}$ $b=5.8^{\circ}$) 
we have obtained recently Na I absorption spectra in directions of 
two close by stars with known distances (Meyer \& Stanimirovic, in preparation). 
As shown in Figure 4, in the direction of star HD 107258, which is 
12 arcmin to the south-east from 3C270 and at a distance of 108 $(+11, -9)$ pc, an
absorption feature was detected at a LSR velocity of 0.55 \kms and with 
a velocity FWHM of 3-4 \kms.  The HI absorption spectrum in Figure 3 shows a feature
at a velocity of $-0.8$ \kms, with a velocity FWHM of 3.2 \kms. 
The two absorption features most likely trace the same interstellar gas.
This suggests that the local HI absorbing gas, seen in Figure 3, is likely to be located 
within 100-120 pc from the Sun. In this direction, 
Na I observations by Lallement et al. (2003) suggest that the boundary 
of the LB, with $N(HI)>10^{20}$ cm$^{-2}$, is at a distance of $\sim130$ pc.
Therefore, the HI cloud in the direction of 3C270 could easily be lying within 
the Local cavity.
This is a cold HI cloud, with the velocity linewidth of  only 3.2 \kms, the estimated
spin temperature of 115 K, and the peak HI column density of $10^{19}$ cm$^{-2}$.
As the 0.55 \kms absorption feature is absent in the Na I spectrum of HD 107238, a 
star which is 14 arcmin to the north-east from 3C270 and at a distance of 192 $(+40, -28)$ pc,
 the spatial extent of the HI absorbing cloud maybe only $\sim0.3$-0.4 pc.
Figure 4 also shows a strong Na I absorption feature at a velocity of $-2.5$ \kms~
seen only in the spectrum against the more distant star and without
a corresponding HI absorption. This (warmer) cloud is likely to be at a distance of 
110--190 pc. Followup observations of several stars around 3C270 are underway 
with the Kitt Peak 3.5m telescope to constrain better cloud distances.

Another especially clear example of a cold HI cloud found inside 
the LB was studied by 
\citet{Meyer06}.  This Leo cloud holds a double record, being the coldest diffuse
cloud with a spin temperature of only 20 K, and at the same time the closest
diffuse cloud with a firm upper distance limit of 45 pc (constrained using 33
stars with known distances). If in pressure
equilibrium with $P=2300$ K cm$^{-3}$,
it has a size of $1.4\times4.9\times0.07$ pc, being more ribbon-like than
spherically symmetric. 
It is worth noting that a thermal pressure close to the standard ISM 
pressure was assumed and this resulted in the pc-scale size of the cold cloud
(a higher thermal pressure would decrease the cloud size).

The Riegel-Crutcher cloud is another cold, highly filamentary cloud, that
appears to be located right at the edge of the LB 
\citep{Crutcher84,McClure-Griffiths06b}.
This cloud extends over $\sim40$ degrees of Galactic longitude ($l=5^{\circ}$ to 355$^{\circ}$)
and $\sim10$ degrees of latitude ($b=-5^{\circ}$ to 5$^{\circ}$). 
It has a distance of $125\pm25$ pc, a size of $80\times20\times(1-5)$ pc, and a
HI spin temperature of 40 K. If filaments are cylindrical with the thickness similar to 
their plane-of-sky width of 0.1 pc,  then they have an average volume density 
of 450 cm$^{-3}$ and a thermal pressure of $2\times10^4$ K cm$^{-3}$.
However, an edge-on ribbon geometry would imply a line-of-sight thickness of 1 pc, 
an average volume density of 46 cm$^{-3}$, and a thermal pressure of
 $2\times10^3$ K cm$^{-3}$.

\section{Comparison with the Cluster of Local Interstellar Clouds}

\begin{table}
\caption{Approximate properties of clouds inside the Local Bubble.}
\label{tab:1}       % Give a unique label
\begin{tabular}{lllll}
\hline\noalign{\smallskip}
Type & Temperature & Density & Size & Ionization state  \\
& (K) & cm$^{-3}$ & (AU) &   \\
\noalign{\smallskip}\hline\noalign{\smallskip}
CNM & $100$ & $40$ & 10$^5$  & neutral\\
TSAS & $\la 100$ & $10^{3-4}$& 10-100 & neutral \\
TSIS &$\ga$10$^{4}$  & $10^{2}$& 1 & ionized (arclets) \\
low-$N(HI)$ & $\la100$ & $10^{2}$& 800-4000 & neutral \\
CLIC & 6800 & 0.2& $\sim10^5$ & partially ionized \\
\noalign{\smallskip}\hline
\end{tabular}
\end{table}

The cluster of local interstellar clouds (CLIC) is a clumpy flow of
warm, low-density material inside the LB, and within 35 pc
from the Sun. Table 1 provides a summary of basic properties of CLIC and
clouds discussed in previous sections.
Further from the Sun,
within 350 pc, the ISM includes low-density, hot gas inside
the LB, but also a range of clouds with cooler temperatures.
For example,
\citet{Welty99} found four neutral, cold ($T\sim100$ K) and dense
($N(HI)\sim10^{20.7}$ cm$^{-2}$ and $n\sim10-15$ cm$^{-3}$) clouds, and
many warm ($T\sim3000$ K), dense ($n\sim15-20$ cm$^{-3}$) clouds.

While CLIC clouds have column densities similar to those of TSAS and 
low-$N(HI)$ clouds, they are significantly warmer and more ionized.
In terms of fractional ionization, CLIC is intermediate 
between the WNM and the warm ionized medium \citep{Dickey04}.
\citet{Muller06} compared the HI column density and the LSR velocity of 
CNM and WNM clouds at high Galactic latitudes (from Heiles \& Troland 2003) 
with those of CLIC clouds within 50 pc of the Sun.
While CLIC clouds have typically a low HI column density ($N(HI)<10^{19}$ cm$^{-2}$), 
their kinematic properties appear similar to those of WNM clouds. 
The central velocity of the majority of low-$N(HI)$ clouds is in the range 
$-17$ to 10 \kms, and the column density range is $3\times10^{17}$ to 
$4\times10^{19}$ cm$^{-2}$. 
In terms of column density and kinematics, low-$N(HI)$ clouds (as well as the
cold Leo cloud and the TSAS clouds in the direction of B1929+10) and CLIC 
occupy a very similar parameter space.
This may suggest that all these clouds belong to the same interstellar flow.

Table 2 provides a summary of exotic clouds inside or at the boundary of the
LB. Clearly, Leo cold clouds are undoubtly inside the LB. 
The low-$N(HI)$ cloud in the direction of 3C270 and the TSAS cloud(s) 
in the direction of B1929+10 are very likely to be located inside the LB.
In addition, several TSIS scintillation screens are clearly located inside the
LB.
The key question is whether warm and cold clouds inside the LB 
are physically related. For example, could CLIC clouds represent warm
envelopes of colder TSAS and low-$N(HI)$ clouds?
And, could scintillation screens represent ionized fronts of CLIC?
As hinted in Section 4, there is evidence for
the existence of large warm envelopes around low-$N(HI)$ clouds.
The existence of cold clouds inside the LB with kinematic properties
similar to those of CLIC rises an exciting possibility that at least some cold
clouds are physically co-located with CLIC. Based on heating and cooling
considerations, the co-existance of warm and
cold phases requires a well-defined range of thermal pressures.
Hence, future distance measurements of low-$N(HI)$ clouds
may allow us to constrain better the thermal pressure inside the LB.

\begin{table}
\caption{Summary of exotic clouds inside or just outside of the Local Bubble.}
\label{tab:1}       % Give a unique label
\begin{tabular}{llll}
\hline\noalign{\smallskip}
Object & Distance & Type & Comment  \\
& (pc) &  &    \\
\noalign{\smallskip}\hline\noalign{\smallskip}
B1929+10 cloud(s) & $<360$ & TSAS, TSIS & cold clouds inside/edge of LB,
several scintillation screens \\
B1133+16 screen & $140$ & TSIS & scintillation screens \citep{Trang07}, no TSAS \\
3C270 cold cloud &$\la100$-120  & low-$N(HI)$& most likely inside LB \\
Leo cloud & $\la45$ & CNM& clearly inside LB \\
R-C cloud & 125 & CNM& at the edge of LB \\
\noalign{\smallskip}\hline
\end{tabular}
\end{table}

 \section{Formation and survival of cold clouds}
 
Is it surprising to find cold clouds inside the LB?
At first glance, we would expect
that cold ``thin'' clouds would be subjected to the heat from the surrounding hot medium
and be eaten away by evaporation very quickly.
However, careful consideration of conductive heat transfer, especially in the
small-scale regime, suggests that such clouds are surprisingly long-lived    
although are most likely undergoing evaporation.
Their typical evaporation timescale is
of order of $10^{6}$ yr (see Stanimirovic \& Heiles 2005 for details; also
\citet{Slavin07}).
If surrounded by large WNM envelopes (as suggested in Section 4), 
and/or magnetic field, the clouds would
be even longer lived, and therefore could be common in the local ISM.

However,
physical mechanisms responsible for the production 
of such clouds in the ISM
in general are not well understood.
There are several potentially interesting theoretical avenues that await a
closer comparison with observations.

 {\it Condensation of WNM into CNM triggered by the collision 
of turbulent flows} 
is capable of producing a large number of small CNM clouds with low column
densities, as seen in simulations by \citet{Audit05} and \citet{Hennebelle07a}.
The CNM clouds produced in simulations are
thermally stable and embedded in large, unstable WNM filaments,
their typical properties are:
$n\sim50$ cm$^{-3}$, $T\sim80$ K, $R\sim0.1$ pc. 
The number of cold clouds, as well as their properties,  depend heavily
on the properties of the underlying turbulent flows. 
Within this scenario, exotic cold clouds inside the LB 
may be remnants of collisions between warmer CLIC clouds. 

{\it General ISM turbulence} 
envisions interstellar clouds as dynamic entities that are constantly
changing in response to the turbulent ``weather'', a picture 
very different from the traditional approach. 
Numerical simulations of turbulent ISM, for example
\citet{Semadeni97} show many clouds
with very small sizes and low column densities ($<10^{19}$ cm$^{-2}$).
These clouds are out of dynamical equilibrium and probably very transient. 
In this scenario, the exotic cold clouds may be a tail-end
of the turbulent spectrum of general ISM clouds.

{\it CNM destruction by shocks} in simulations by \citet{Nakamura06}
can also produce a spray of small HI `shreds' that could be related to
low-$N(HI)$ clouds.  These shreds are expected to have large aspect ratios, up to 2000.
As several successive supernova explosions have been postulated to be
responsible for the formation of
 the LB \citep{Breitschwerdt06},
it would not be hard to imagine shock interactions with 
clouds inside the LB and the production of exotic cloudlets. 
Similarly,
Breitschwerdt \& de Avillez (2006) suggest that the interaction between
shells that bound bubbles caused by supernova explosions can drive
Rayleigh-Taylor instability and the formation of cloudlets.

\section{Conclusions}

We have summarized evidence for the existence of at least several exotic 
clouds inside or just outside the LB.
The importance of these clouds is twofold. Firstly, they provide extreme
examples of the interstellar clouds that the Sun may encounter
over a timescale of several million years. As shown by \citet{Muller06},
highly dense and cold clouds would have a significant effect on the size of the
heliosphere. Secondly, studies of different phases of the very local ISM
can shed some light on the properties and the formation mechanism of the LB.

The cold exotic clouds (TSAS and low-$N(HI)$ clouds) 
have AU-scale sizes and low HI column densities. 
While TSAS has high HI volume density and thermal pressure, 
low-$N(HI)$ clouds are in this regard similar to traditional CNM clouds.
While significantly colder and less ionized than CLIC, the exotic
(cold) clouds agree well with CLIC clouds in terms of HI column density and kinematics.
This may suggest the physical co-existence of different populations, however
distances of most of low-$N(HI)$ clouds are unknown.
If physically co-located, these clouds would imply a well-defined range of 
thermal pressures inside the LB. 
Future distance constrains of cold clouds are essential to explore this avenue.  

While most likely undergoing thermal evaporation, cold exotic clouds
inside the LB could last for a long time. Therefore, it is
not surprising to encounter them frequently.
However, their production mechanisms are still not understood.
Potential avenues highlight the importance of dynamical triggering
of phase conversion (through collision of warm clouds or
interaction of shocks with warm clouds), and interstellar turbulence.

\begin{acknowledgements}
It is a great pleasure to acknowledge my collaborators Joel Weisberg, Carl
Heiles, Dave Meyer, and Evan De Back, without whom this work would not have 
been so much fun. I would also like to thank the organizers for
inviting me to participate in this highly stimulating conference.
Finally, this article has greatly benefited from the editors' insightful comments. 
\end{acknowledgements}

% BibTeX users:
%\newcommand{\bibfont}{\footnotesize}
%\bibliographystyle{SSRv}
%\bibliography{/d/leffe/sstanimi/rattler/Thesis/Thesisdir/bib/astro-mnemonic,/d/leffe/sstanimi/rattler/Thesis/Thesisdir/bib/myref}   % name your BibTeX data base

%% Non-BibTeX users please use
%\begin{thebibliography}{}
%%
%% and use \bibitem to create references, as follows:
%\end{thebibliography}

\end{document}